\documentclass{article}
\usepackage{amssymb}

\title{A Structure behind Primitive Chaos}

\author{%
Yoshihito Ogasawara\thanks{E-mail address: ogasawara@aoni.waseda.jp}\\
Faculty of Science and Engineering, Waseda University, Japan}

\begin{document}
%\maketitle

\renewcommand\thefootnote{*}
\begin{center}
{\LARGE A Structure behind Primitive Chaos}\\~\\
{\large Yoshihito Ogasawara\footnote{E-mail address: ogasawara@aoni.waseda.jp}}\\
{\large Faculty of Science and Engineering, Waseda University, Japan}\\
\end{center}

\noindent{\bf Abstract}

Recently, a new concept, primitive chaos, was proposed, as a concept closely related to the fundamental problems of sciences themselves such as determinism, causality, free will, predictability, and irreversibility  [J. Phys. Soc. Jpn. {\bf 79}, 15002 (2010)]. This letter reveals a structure hidden behind the primitive chaos; under some conditions, a new primitive chaos is constructed from the original primitive chaos, this procedure can be repeated, and the hierarchic structure of the primitive chaos is obtained. This implies such a picture that new events and causality is constructed from the old ones, with the aid of the concept of a coarse graining. As an application of this structure, interesting facts are revealed for the essential condition of the primitive chaos and for the chaotic behaviors.\\


%\newpage
\section{Introduction}

Recently, a new concept, primitive chaos, was proposed, as a concept closely related to the fundamental problems of sciences themselves such as determinism, causality, free will, predictability, and irreversibility \cite{O2010,O2014}.\\

%\vspace{2mm}
{\noindent\bf Definition 1.}~{\it If a set $X$, the family of nonempty subsets of $X$, $\{X_\lambda,~\lambda\in\Lambda\}$, and the family of maps, $\{f_{X_\lambda}:X_\lambda\to X,~\lambda\in\Lambda\}$, satisfy the following property (P), $(X,~\{X_\lambda,~\lambda\in\Lambda\},~\{f_{X_\lambda},~\lambda\in\Lambda\})$ is called a primitive chaos.
\begin{itemize}
\item[(P)] For any infinite sequence $\omega_0,~\omega_1,~\omega_2,\ldots$, there exists an initial point $x_0\in\omega_0$ such that
\begin{eqnarray}
f_{\omega_0}(x_0)\in\omega_1,~f_{\omega_1}(f_{\omega_0}(x_0))\in\omega_2,\ldots,\label{P}
\end{eqnarray} where each $\omega_i$ is an element of the family $\{X_\lambda,~\lambda\in\Lambda\}$.
\end{itemize}}
\vspace{4mm}

\noindent In the primitive chaos, each set $X_\lambda$ implies an event or a selection, each sequence $(\omega_n)_{n=0}^\infty$ implies a time series, and each map $f_{X_\lambda}$ implies a law or causality \cite{O2010,O2014}. 

Under natural conditions, the primitive chaos leads to 
the characteristic properties of the conventional chaos \cite{O2012,RLD}, such as the existence of a nonperiodic orbit, the existence of the periodic point whose prime period is $n$ for any $n\in\mathbb{N}$, the existence of a dense orbit, the density of periodic points, sensitive dependence on initial conditions, and topological transitivity. In this sense, this primitive chaos is literally a primitive chaos.

Then, the following proposition was proposed \cite{O2010}, which exhibits sufficient conditions for the guarantee of existence of the primitive chaos. Here, a topological space $X$ is said to be countably compact, provided that every countable open cover of $X$ has a finite subcover; thus, any compact space is countably compact.\\

%\vspace{2mm}
{\noindent\bf Proposition 1.}~{\it If $X$ is a countably compact space, $\{X_\lambda,~\lambda\in\Lambda\}$ is a family of nonempty closed subsets of $X$, and $\{f_{X_\lambda},~\lambda\in\Lambda\}$ is a family of continuous surjections, then they satisfy the property (P).}~\\
%\vspace{2mm}

However, the condition of $\{f_{X_\lambda},~\lambda\in\Lambda\}$ seems to be too artificial or strong \cite{O2010,SBN}, and thus we explored conditions for the guarantee of existence of such maps from a topological viewpoint. Then, we attained two characteristic concepts, nondegenerate Peano continuum and Cantor set, along with the concepts of hierarchy, coarse graining, self-similarity, and logic \cite{O2010,O2014,O2011}.

A nondegenerate set means a set consisting of more than one point. A Peano continuum is a locally connected continuum, and a continuum is a nonempty compact connected metric space. The nondegenerate Peano continuum is a general or universal concept such that it has many examples such as all arcs, all $n$-cells, all $n$-spheres, all toruses, all solid toruses, all trees, all graphs, all nondegenerate dendrites, and all Hilbert cubes \cite{SBN}. 

A Cantor set is a space homeomorphic to the Cantor middle-third set, and it is known that a space is a Cantor set if and only if it is a zero-dimensional perfect compact metrizable space \cite{SBN}. A topological space is zero-dimensional provided that there is a base for its topology such that each element of the base is clopen (closed and open), and a topological space is perfect provided that it contains no isolated points. The Cantor set is also a general or universal concept, quite differently from the special set, the Cantor middle-third set \cite{O2014,SBN}.  

 If $X$ is the nondegenerate Peano continuum or the Cantor set, $X$ guarantees not only the existence of the primitive chaos, but also the existence of its infinite variety \cite{O2010,O2014}. Then, these results seem to be an answer of the question ``Why we are surrounded by diverse chaotic behaviors?"\cite{O2014,O2012}. 

In addition, we can see the contrast of these concepts because the nondegenerate Peano continuum is characterized by its continuum and the Cantor set is characterized by its zero-dimensionality. This contrast reminds us of two contrast aspects of matter from a macroscopic viewpoint and a microscopic viewpoint. Then, the concepts of continuity (continuum) and discreteness (zero-dimensionality) seem to be our intrinsic concepts for the method of recognizing phenomena \cite{O2014}.

Then, we obtained the concept of hierarchic structure of the primitive chaos from the relation of whole and part \cite{O2014,O2011}. If $X$ is a nondegenerate Peano continuum (or a Cantor set), it has infinitely variety of the primitive chaos such that for each primitive chaos $(X,~\{X_\lambda,~\lambda\in\Lambda\},~\{f_{X_\lambda},~\lambda\in\Lambda\})$, each set $X_\lambda$ is also a nondegenerate Peano continuum (or a Cantor set) as a subspace of the whole space $X$. Therefore, each set $X_\lambda$ itself has infinitely variety of the primitive chaos. This procedure can be repeated, and we can see the hierarchic structure of the primitive chaos.  

In this study, we obtain another hierarchic structure of the primitive chaos, by focusing on the time series $(\omega_n)_{n=0}^\infty$ of the property (P). By exploring a space consisting of the time series, we see the emergence of events and causality of new primitive chaos from the original ones. 

\section{Hierarchic Structure of Primitive Chaos}

At first, let us suppose that the primitive chaos $(X,~\{X_\lambda,~\lambda\in\Lambda\},~\{f_{X_\lambda},~\lambda\in\Lambda\})$ satisfies the following conditions (\ref{cup})-(\ref{ch'}) which play an important role in leading the primitive chaos to the properties of the conventional chaos \cite{O2012}.
\begin{eqnarray}
X=\bigcup_{\lambda\in\Lambda}X_\lambda.\label{cup} 
\end{eqnarray}
\begin{eqnarray}
\Lambda=\{1,\ldots,m\},~m\ge 2;\label{l}
\end{eqnarray}
that is, $\Lambda$ is a nondegenerate finite set.
\begin{eqnarray}
(X,d);\label{met}
\end{eqnarray}
that is, $X$ is a metric space equipped with metric $d$.
\begin{eqnarray}
\sup_{\omega_0,\ldots,\omega_n\in\{X_{\lambda},~\lambda\in\Lambda\}}dia_d\,f_{X_{\omega_0}}^{-1}(\cdots f_{X_{\omega_n}}^{-1}(X)\cdots)\to 0,\cr
(n\to\infty), \label{ch'}
\end{eqnarray}
where $dia_d\,A$ denotes the diameter of $A$ for the metric $d$.

Then, let us consider the set of all the sequences in the property (P) of the primitive chaos,
\begin{eqnarray}
\Sigma=\{(\omega_n)_{n=0}^\infty|~\omega_n\in\{X_1,\ldots,X_m\}\}.
\end{eqnarray}
Defining the discrete topology for the set $\{X_1,\ldots,X_m\}$, from the relation
\begin{eqnarray}
\Sigma=\prod_{n=0}^\infty Y_n~(Y_n\equiv \{X_1,\ldots,X_m\}),
\end{eqnarray}
we can introduce the product topology for $\Sigma$; that is, we can consider a sequence space \cite{RLD}. 

Since the space $\{X_1,\ldots,X_m\}$ is compact, the space $\Sigma$ is compact by the Tychonoff's theorem. Since the space $\{X_1,\ldots,X_m\}$ is metrizable, the space $\Sigma$ is also metrizable. For a point $(\omega_n)_{n=0}^\infty\in\Sigma$ and an open set $U$ containing $(\omega_n)_{n=0}^\infty$, by the definition of the product topology, there exists a number $N$ such that 
\begin{eqnarray}
(\omega_n)_{n=0}^\infty\in \{\omega_0\}\times\cdots\times\{\omega_N\}\times\prod_{n=N+1}^\infty Y_n\subset U,
\end{eqnarray}
and thus the space $\Sigma$ is perfect because $m\ge 2$. In addition, since the set
\begin{eqnarray}
\bigcup_{i=0}^N Y_0\times\cdots\times Y_{i-1}\times(Y_{i}-\{\omega_i\})\times\prod_{n=i+1}^\infty Y_n
\end{eqnarray}
is an open subset of $\Sigma$, the set $\{\omega_0\}\times\cdots\times\{\omega_N\}\times\prod_{n=N+1}^\infty Y_n$ is a clopen subset of $\Sigma$, and thus the space $\Sigma$ is zero-dimensional. Namely, the space $\Sigma$ is a Cantor set, and thus it itself can infinitely generate the primitive chaos. 

Next, let us consider the shift map
\begin{eqnarray}
\sigma:\Sigma\to\Sigma,~(\omega_n)_{n=0}^\infty\mapsto(\omega_n')_{n=0}^\infty~(\omega_n'=\omega_{n+1}),
\end{eqnarray}
that is, the symbolic dynamics \cite{RLD}. This map is obviously a surjection, and from the relation
\begin{eqnarray}
&&\sigma(\{\omega_0\}\times\cdots\times\{\omega_N\}\times\prod_{n=N+1}^\infty Y_n)\cr
&=&
\{\omega_1\}\times\cdots\times\{\omega_{N}\}\times\prod_{n=N+1}^\infty Y_n
\end{eqnarray}
for any $N$, the map $\sigma$ is continuous. 

Let us explore the relation between this map $\sigma$ and the family
$\{f_{X_1},\ldots,f_{X_m}\}$ of the primitive chaos. At first, note that the condition (\ref{ch'}) leads to the uniqueness of the initial point $x_0$ of the property (P) \cite{O2012}. Therefore, we can consider the map
\begin{eqnarray}
g:\Sigma\to X,~(\omega_n)_{n=0}^\infty\mapsto x_0,
\end{eqnarray}
where $x_0$ is the unique initial point of the property (P) for the given sequence $(\omega_n)_{n=0}^\infty$. From the condition (\ref{cup}), for any $x_0\in X$, there exists a sequence $(\omega_n)_{n=0}^\infty$ satisfying the condition (\ref{P}), and thus the map $g$ is a surjection.

In addition, the map $g$ is continuous. Let $U$ be the open subset of $X$, which contains the point $g((\omega_n)_{n=0}^\infty)$ in $X$. From the condition (\ref{ch'}), there exists a number $N$ such that 
\begin{eqnarray}
f_{X_{\omega_{0}}}^{-1}(\cdots f_{X_{\omega_{N}}}^{-1}(X)\cdots)\subset U.
\end{eqnarray}
Then, since the relation
\begin{eqnarray}
g((\omega_n)_{n=0}^\infty)&\in&g(\{\omega_0\}\times\cdots\times\{\omega_N\}\times\prod_{n=N+1}^\infty Y_n)\cr
&=&f_{X_{\omega_{0}}}^{-1}(\cdots f_{X_{\omega_{N}}}^{-1}(X)\cdots)
\end{eqnarray}
is satisfied, the map $g$ is continuous.

Furthermore, for each $j$, the relation
\begin{eqnarray}
g\circ\sigma|_{g^{-1}(X_j)}=f_{X_j}\circ g|_{g^{-1}(X_j)}\label{fg}
\end{eqnarray}
is obtained. For each $(\omega_n)_{n=0}^\infty\in g^{-1}(X_j)$ and any $N$, 
\begin{eqnarray}
g((\omega_n)_{n=0}^\infty)\in f_{X_j}^{-1}(f_{X_{\omega_1}}^{-1}(\cdots f_{X_{\omega_{N}}}^{-1}(X)\cdots)),
\end{eqnarray}
and thus 
\begin{eqnarray}
f_{X_j}(g((\omega_n)_{n=0}^\infty))
\in f_{X_{\omega_{1}}}^{-1}(\cdots f_{X_{\omega_{N}}}^{-1}(X)\cdots).
\end{eqnarray}
Therefore,
\begin{eqnarray}
f_{X_j}(g((\omega_n)_{n=0}^\infty))=g((\omega_n)_{n=1}^\infty)=g(\sigma((\omega_n)_{n=0}^\infty)).
\end{eqnarray}
Consequently, we can now see such a situation that by the use of the continuous surjection $g:\Sigma\to X$, the family
$\{f_{X_1},\ldots,f_{X_m}\}$ of the primitive chaos is correlated with the shift map $\sigma:\Sigma\to\Sigma$. 

Let us further explore this relation. At first, recall the concept of a decomposition space, which can be recognized as a coarse graining \cite{SBN,O2011,YO}.

~\\
%\vspace{2mm}
{\noindent\bf Definition 2.} {\it For a topological space $(Y,\tau)$ and a partition $\mathcal{D}$ of $Y$, 
\begin{eqnarray}
\{\mathcal{U}\subset\mathcal{D}|~\bigcup_{U\in\mathcal{U}}U\in\tau\}[=\tau(\mathcal{D})]
\end{eqnarray}
is called a decomposition topology, and $(\mathcal{D},\tau(\mathcal{D}))$ is called a decomposition space of $(Y,\tau)$.}
~\\
%\vspace{2mm}

Then, let us consider the partition of the sequence space $\Sigma$ as a quotient set by the equivalence relation
\begin{eqnarray}
(\omega_n)_{n=0}^\infty\sim(\omega'_n)_{n=0}^\infty\Leftrightarrow g((\omega_n)_{n=0}^\infty)=g((\omega'_n)_{n=0}^\infty);
\end{eqnarray}
recall that the point $g((\omega_n)_{n=0}^\infty)$ is the unique initial point describing the sequence $(\omega_n)_{n=0}^\infty$ in the property (P). Then, since $X$ is a Hausdorff space, $\Sigma$ is compact, and $g:\Sigma\to X$ is a continuous surjection, the map
\begin{eqnarray}
h:\mathcal{D}_g\to X,~g^{-1}(x)\mapsto x
\end{eqnarray}
is a homeomorphism, where $\mathcal{D}_g=\{g^{-1}(x)|~x\in X\}$ \cite{O2011,YO}. Accordingly, if the original space $X$ is a nondegenerate Peano continuum or a Cantor set, the decomposition space $\mathcal{D}_g$ of the sequence space $\Sigma$ is so, and thus it itself can infinitely generate the primitive chaos. 

Then, we can now obtain a new primitive chaos
$$
(\mathcal{D}_g,\{h^{-1}(X_1),\ldots,h^{-1}(X_m)\},\{f_{h^{-1}(X_1)},\ldots,f_{h^{-1}(X_m)}\})
$$
from the original primitive chaos $(X,\{X_1,\ldots,X_m\},\{f_{X_1},\ldots,f_{X_m}\})$, where
\begin{eqnarray}
f_{h^{-1}(X_j)}:h^{-1}(X_j)\to\mathcal{D}_g, D\mapsto h^{-1}\circ f_{X_j}\circ h(D)\label{fxi}
\end{eqnarray}
for each $j$. For any sequence $(\omega_n)_{n=0}^\infty$ such that each sequence $\omega_n$ is an element of the family $\{h^{-1}(X_1),\ldots,h^{-1}(X_m)\}$, there exists an initial point $x_0\in h(\omega_0)$ such that 
\begin{eqnarray}
f_{h(\omega_0)}(x_0)\in h(\omega_1),~f_{h(\omega_1)}(f_{h(\omega_0)}(x_0))\in h(\omega_2),\ldots,
\end{eqnarray}
and thus the relation
\begin{eqnarray}
h^{-1}(x_0)\in \omega_0,\cr
f_{\omega_0}(h^{-1}(x_0))=h^{-1}(f_{h(\omega_0)}(x_0))\in \omega_1,\cr
f_{\omega_1}(f_{\omega_0}(h^{-1}(x_0)))=h^{-1}(f_{h(\omega_1)}(f_{h(\omega_0)}(x_0)))\in \omega_2,\cr
\vdots~~~~\,
\end{eqnarray}
is satisfied. Accordingly, the property (P) is satisfied for 
$$
(\mathcal{D}_g,\{h^{-1}(X_1),\ldots,h^{-1}(X_m)\},\{f_{h^{-1}(X_1)},\ldots,f_{h^{-1}(X_m)}\}); 
$$
that is, it is a new primitive chaos obtained from the original one.

In addition, from the condition (\ref{cup}), the relation
\begin{eqnarray}
h^{-1}(X_1)\cup\cdots\cup h^{-1}(X_m)=\mathcal{D}_g
\end{eqnarray}
is satisfied. Then, from the original metric $d$ of $X$, the metric of $\mathcal{D}_g$,
\begin{eqnarray}
\rho(D,D')=d(h(D),h(D')),\label{rho}
\end{eqnarray}
can be introduced such that the metric space $(\mathcal{D}_g,\rho)$ is homeomorphic to the decomposition space $(\mathcal{D}_g,\tau(\mathcal{D}_g))$ \cite{CSF2007}.

Then, the relation
\begin{eqnarray}
\sup_{\omega_0,\ldots,\omega_n\in\{h^{-1}(X_1),\ldots,h^{-1}(X_m)\}}dia_\rho\,f_{\omega_0}^{-1}(\cdots f_{\omega_n}^{-1}(\mathcal{D}_g)\cdots)\to 0,\cr
(n\to\infty)\label{ch''}
\end{eqnarray}
is satisfied. For $\omega_0,\ldots,\omega_n\in\{h^{-1}(X_1),\ldots,h^{-1}(X_m)\}$, from the relation
\begin{eqnarray}
f_{\omega_n}\circ\cdots\circ f_{\omega_0}=h^{-1}\circ f_{h(\omega_n)}\circ\cdots\circ f_{h(\omega_0)}\circ h,
\end{eqnarray}
the relation
\begin{eqnarray}
f_{\omega_0}^{-1}(\cdots f_{\omega_n}^{-1}(\mathcal{D}_g)\cdots)
=h^{-1}(f_{h(\omega_0)}^{-1}(\cdots f_{h(\omega_n)}^{-1}(X)\cdots))
\end{eqnarray}
is satisfied. Then, by the definition (\ref{rho}), the relation
\begin{eqnarray}
&&dia_\rho\,f_{\omega_0}^{-1}(\cdots f_{\omega_n}^{-1}(\mathcal{D}_g)\cdots)\cr
&=&dia_d\,f_{h(\omega_0)}^{-1}(\cdots f_{h(\omega_n)}^{-1}(X)\cdots)
\end{eqnarray}
is satisfied; that is, the relation (\ref{ch''}) is obtained.

Consequently, we can obtain the new symbolic dynamics with respect to the new primitive chaos, and its decomposition space. Since this procedure can be repeated, we can see the new hierarchic structure of the primitive chaos.

\section{Properties}

Let us further explore the structure given in the previous section. At first, we can see the relation
\begin{eqnarray}
\sigma(g^{-1}(x))\subset f_{h^{-1}(X_j)}(g^{-1}(x))\label{sigma}
\end{eqnarray}
for each $j$ and any $x\in X_j$, as the relation of the maps $\sigma:\Sigma\to\Sigma$ and $f_{h^{-1}(X_j)}:h^{-1}(X_j)\to \mathcal{D}_g$. For each $(\omega_n)_{n=0}^\infty\in\sigma(g^{-1}(x))$, from the relation (\ref{fg}), the relation
\begin{eqnarray}
g((\omega_n)_{n=0}^\infty)&\in&g(\sigma(g^{-1}(x)))\cr
&=&f_{X_j}(g(g^{-1}(x)))\cr
&=&\{f_{X_j}(x)\}
\end{eqnarray}
is satisfied. Therefore, the relation
\begin{eqnarray}
(\omega_n)_{n=0}^\infty&\in&g^{-1}(f_{X_j}(x))\cr
&=&h^{-1}(f_{X_j}(x))\cr
&=&f_{h^{-1}(X_j)}(h^{-1}(x))\cr
&=&f_{h^{-1}(X_j)}(g^{-1}(x))
\end{eqnarray}
is satisfied; that is, the relation (\ref{sigma}) is obtained.

Then, we can see the fact that each map $f_{X_j}$ is a surjection, although this condition was unrequired. For each $j$, since the relation
\begin{eqnarray}
g^{-1}(X_j)=\{X_j\}\times\prod_{n=1}^\infty Y_n
\end{eqnarray}
is satisfied, from the relation (\ref{fg}), the relation
\begin{eqnarray}
f_{X_j}(X_j)=f_{X_j}(g(g^{-1}(X_j)))=g(\sigma(g^{-1}(X_j)))=X
\end{eqnarray}
is obtained; that is, $f_{X_j}$ is a surjection.

In addition, if each set $X_j$ is closed as required in Proposition 1, each map $f_{X_j}$ is continuous, although this condition was also unrequired. Let a continuous surjection $q:\Sigma\to\mathcal{D}_g$ be defined by
\begin{eqnarray}
q=h^{-1}\circ g.
\end{eqnarray}
From the relation (\ref{fg}), the relation
\begin{eqnarray}
f_{h^{-1}(X_j)}\circ q|_{g^{-1}(X_j)}&=&h^{-1}\circ f_{X_j}\circ h\circ q|_{g^{-1}(X_j)}\cr
&=&h^{-1}\circ f_{X_j}\circ g|_{g^{-1}(X_j)}\cr
&=&h^{-1}\circ g\circ\sigma|_{g^{-1}(X_j)}\cr
&=&q\circ\sigma|_{g^{-1}(X_j)}
\end{eqnarray}
is satisfied. Then, for a closed subset $\mathcal{F}$ of $\mathcal{D}_g$, from the relation
\begin{eqnarray}
q^{-1}(f_{h^{-1}(X_j)}^{-1}(\mathcal{F}))&=&g^{-1}(h(f_{h^{-1}(X_j)}^{-1}(\mathcal{F})))\cr
&\subset&g^{-1}(h(h^{-1}(X_j))))\cr
&=&g^{-1}(X_j),
\end{eqnarray}
the subset of the compact space $\Sigma$,
\begin{eqnarray}
q^{-1}(f_{h^{-1}(X_j)}^{-1}(\mathcal{F}))=\sigma^{-1}(q^{-1}(\mathcal{F}))\cap g^{-1}(X_j),
\end{eqnarray}
is closed, and thus compact. Then, the subset of the Hausdorff space $\mathcal{D}_g$,
\begin{eqnarray}
f_{h^{-1}(X_j)}^{-1}(\mathcal{F})=q(\sigma^{-1}(q^{-1}(\mathcal{F}))\cap g^{-1}(X_j)),
\end{eqnarray}
is compact, and thus closed. That is, the map $f_{h^{-1}(X_j)}$ is continuous, and thus the map $f_{X_j}$ is continuous by the definition (\ref{fxi}).

These results are interesting. We obtained Proposition 1 by exploring sufficient conditions for the guarantee of existence of the primitive chaos. Then, the condition of $\{f_{X_\lambda},~\lambda\in\Lambda\}$ (that is, each of them is a continuous surjection) seemed to be too artificial or strong. However, we can now see that the condition is not only a sufficient condition, but also a necessary condition if each $X_j$ is closed and the conditions (\ref{cup})-(\ref{ch'}) are satisfied.

Furthermore, let us consider the following condition for the original primitive chaos,
\begin{eqnarray}
x\in X_{j}\cap X_{j'}\Rightarrow f_{X_{j}}(x)=f_{X_{j'}}(x),\label{ya}
\end{eqnarray}
in addition to the conditions (\ref{cup})-(\ref{ch'}). From the condition (\ref{cup}), the unified map
\begin{eqnarray}
f:X\to X,~x\mapsto f_{X_j}(x)~(x\in X_j)
\end{eqnarray}
can be defined, and $f$ has the chaotic properties such as the existence of a dense orbit, the density of periodic points, and topological transitivity \cite{O2012}.

As described in the previous section, the new primitive chaos 
$$
(\mathcal{D}_g,\{h^{-1}(X_1),\ldots,h^{-1}(X_m)\},\{f_{h^{-1}(X_1)},\ldots,f_{h^{-1}(X_m)}\})
$$
is constructed from the original one $(X,\{X_1,\ldots,X_m\},\{f_{X_1},\ldots,f_{X_m}\})$. 
Then, the condition corresponding to the condition (\ref{ya}),
\begin{eqnarray}
&&D\in h^{-1}(X_{j})\cap h^{-1}(X_{j'})\cr
&\Rightarrow&f_{h^{-1}(X_{j})}(D)=f_{h^{-1}(X_{j'})}(D),\label{ya'}
\end{eqnarray}
is satisfied. As a matter of fact, for $D\in h^{-1}(X_{j})\cap h^{-1}(X_{j'})$, since $h(D)\in X_{j}\cap X_{j'}$, from the condition (\ref{ya}), the relation
\begin{eqnarray}
f_{h^{-1}(X_{j})}(D)&=&h^{-1}(f_{X_{j}}(h(D)))\cr
&=&h^{-1}(f_{X_{j'}}(h(D)))\cr
&=&f_{h^{-1}(X_{j'})}(D)
\end{eqnarray}
is satisfied. Accordingly, the unified map
\begin{eqnarray*}
\tilde{f}:\mathcal{D}_g\to \mathcal{D}_g,~D\mapsto f_{h^{-1}(X_{j})}(D)~(D\in h^{-1}(X_{j}))
\end{eqnarray*}
can be defined, and it has the properties such as the existence of a dense orbit, the density of periodic points, and topological transitivity. Since each map $f_{h^{-1}(X_{j})}$ is a surjection, the map $\tilde{f}$ is also a surjection. 

Here, note that the relation
\begin{eqnarray}
\tilde{f}=h^{-1}\circ f\circ h\label{tilf}
\end{eqnarray}
is satisfied as the relation of the unified maps. As a matter of fact, for $D\in h^{-1}(X_{j})$, since $h(D)\in X_{j}$, the relation
\begin{eqnarray}
\tilde{f}(D)&=&f_{h^{-1}(X_{j})}(D)\cr
&=&h^{-1}(f_{X_{j}}(h(D)))\cr
&=&h^{-1}(f(h(D)))
\end{eqnarray}
is satisfied.

Then, the map $\tilde{f}$ is continuous, regardless of the closeness of each $X_{j}$. From the relation (\ref{fg}), the relation
\begin{eqnarray}
g\circ\sigma={f}\circ g,\label{fg'}
\end{eqnarray}
is satisfied. Then, from the relation (\ref{tilf}), the relation
\begin{eqnarray}
\tilde{f}\circ q&=&h^{-1}\circ f\circ h\circ h^{-1}\circ g\cr
&=&h^{-1}\circ f\circ g\cr
&=&q\circ\sigma
\end{eqnarray}
is satisfied, and thus for a closed subset $\mathcal{F}$ of $\mathcal{D}_g$,
\begin{eqnarray}
\tilde{f}^{-1}(\mathcal{F})=q(\sigma^{-1}(q^{-1}(\mathcal{F})))
\end{eqnarray}
is closed; that is, the map $\tilde{f}$ is continuous. Therefore, from the condition (\ref{tilf}), the map $f$ is a continuous surjection, even though such a condition of $f$ was not supposed.

Lastly, let us consider the following conditions (\ref{2}) and (\ref{N}) for the original primitive chaos in addition to the conditions (\ref{cup})-(\ref{ch'}) and (\ref{ya}).
\begin{eqnarray}
|X|\ge 2;\label{2}
\end{eqnarray}
that is, $X$ is nondegenerate.
\begin{eqnarray}
x\in X_{j}\cap X_{j'}\Rightarrow \exists I~{\rm s.t.}~f^i(x)\notin X_{j}\cap X_{j'},~\forall i\ge I. \label{N}
\end{eqnarray}
Then, the unified map $f$ has the chaotic properties such as nonperiodic orbits, the periodic point whose prime period is $n$ for any $n\in\mathbb{N}$, and sensitive dependence on initial conditions \cite{O2012}. 

Here, the decomposition space $\mathcal{D}_g$ is also nondegenerate. Then, the condition corresponding to the condition (\ref{N}),
\begin{eqnarray}
D\in h^{-1}(X_{j})\cap h^{-1}(X_{j'})\cr
\Rightarrow \exists I~{\rm s.t.}~\tilde{f}^i(D)\notin h^{-1}(X_{j})\cap h^{-1}(X_{j'}),~\forall i\ge I, \label{NN}
\end{eqnarray}
is satisfied. For $D\in h^{-1}(X_{j})\cap h^{-1}(X_{j'})$, from the relation $h(D)\in X_{j}\cap X_{j'}$, there exists $I$ such that ${f}^i(h(D))\notin X_{j}\cap X_{j'}$ for any $i\ge I \label{N'}$. Then, from the relation
\begin{eqnarray}
h(\tilde{f}^i(D))={f}^i(h(D))\notin X_{j}\cap X_{j'},
\end{eqnarray}
the relation
\begin{eqnarray}
\tilde{f}^i(D)\notin h^{-1}(X_{j}\cap X_{j'})=h^{-1}(X_{j})\cap h^{-1}(X_{j'})
\end{eqnarray}
is satisfied; that is, the condition (\ref{NN}) is obtained.  Accordingly, the unified map $\tilde{f}$ also has nonperiodic orbits, the periodic point whose prime period is $n$ for any $n\in\mathbb{N}$, and sensitive dependence on initial conditions.

\section{Conclusions}

As the concept of primitive chaos is a starting point, a fertile structure hidden behind it is revealed. 

Under the conditions (\ref{cup})-(\ref{ch'}), the hierarchic structure of the primitive chaos is constructed. The space $\Sigma$ consisting of all the sequences of the family $\{X_1,\ldots,X_m\}$ of the primitive chaos is a Cantor set. The shift map $\sigma:\Sigma \to \Sigma$ is a continuous surjection, and the map $g:\Sigma \to X,~(\omega_n)_{n=0}^\infty\mapsto x_0$ is also a continuous surjection, where $x_0$ is the unique initial point of the property (P) for the given sequence $(\omega_n)_{n=0}^\infty$. Then, the homeomorphism $h:\mathcal{D}_g\to X,~g^{-1}(x)\mapsto x$ is obtained, where $\mathcal{D}_g=\{g^{-1}(x),~x\in X\}$ is the decomposition space of $\Sigma$. By using $h$, we obtain the new primitive chaos 
$$
(\mathcal{D}_g,\{h^{-1}(X_1),\ldots,h^{-1}(X_m)\},\{f_{h^{-1}(X_1)},\ldots,f_{h^{-1}(X_m)}\}).
$$
Here, each map $f_{h^{-1}(X_j)}:h^{-1}(X_j)\to\mathcal{D}_g$ is defined by $f_{h^{-1}(X_j)}(D)=h^{-1}\circ f_{X_j}\circ h(D)$, and the relation $\sigma(g^{-1}(x))\subset f_{h^{-1}(X_j)}(g^{-1}(x))$ is satisfied. Since the primitive chaos $(\mathcal{D}_g,\{h^{-1}(X_1),\ldots,h^{-1}(X_m)\},\{f_{h^{-1}(X_1)},\ldots,f_{h^{-1}(X_m)}\})$ also satisfies the conditions corresponding to the conditions (\ref{cup})-(\ref{ch'}), this procedure can be repeated, and thus the hierarchic structure of the primitive chaos is obtained. 

This structure leads to interesting results. Although surjectivity and continuity were unrequired for each map $f_{X_j}$, we can obtain the fact that it is a continuous surjection if each set $X_j$ is closed as required in Proposition 1. We obtained Proposition 1 by exploring sufficient conditions for the guarantee of existence of the primitive chaos. Then, the condition of $\{f_{X_\lambda},~\lambda\in\Lambda\}$ (that is, each of them is a continuous surjection) seemed to be too artificial or strong. However, we can see that the condition is not only a sufficient condition, but also a necessary condition if each $X_j$ is closed and the conditions (\ref{cup})-(\ref{ch'}) are satisfied.

Under the condition (\ref{ya}) in addition to the conditions (\ref{cup})-(\ref{ch'}), the unified map $f:X\to X,~x\mapsto f_{X_j}(x)~(x\in X_j)$ is obtained such that it has the chaotic properties such as the existence of a dense orbit, the density of periodic points, and topological transitivity. Then, the unified map $\tilde{f}:\mathcal{D}_g\to \mathcal{D}_g,~D\mapsto f_{h^{-1}(X_{j})}(D)~(D\in h^{-1}(X_{j}))$ is also obtained such that $\tilde{f}=h^{-1}\circ f\circ h$, and it also has such chaotic properties. The map $\tilde{f}$ is a continuous surjection regardless of the closeness of each set $X_j$, and the map $f$ is so, even though such a condition of $f$ was not supposed.

Under the conditions (\ref{2}) and (\ref{N}) in addition to the conditions (\ref{cup})-(\ref{ch'}) and (\ref{ya}), the unified map $f$ has nonperiodic orbits, the periodic point whose prime period is $n$ for any $n\in\mathbb{N}$, and sensitive dependence on initial conditions, and the unified map $\tilde{f}$ also has them. That is, as a primitive chaos is a starting point, a fertile structure is constructed, and it leads to interesting results.

Here, recall that, in the context of the primitive chaos, each set $X_i$, each sequence $(\omega_n)_{n=0}^\infty$, each map $f_{X_i}$, and a decomposition space imply an event, a time series, causality, and a coarse graining, respectively. Accordingly, the sequence space $\Sigma$ implies a space consisting of all the time series, and the decomposition space $\mathcal{D}_g$ implies its coarse graining. Then, we can see such a picture that by the coarse graining, a new primitive chaos is obtained; that is, a new event $h^{-1}(X_i)$ and a new causality $f_{h^{-1}(X_i)}$ are constructed from the original ones. Each event $h^{-1}(X_i)$ is the part of a coarse graining of a space consisting of all time series of the events of the original primitive chaos, and the causality $f_{h^{-1}(X_i)}$ is defined on it.

These consequences make us expect the possibility of a new field extending from the primitive chaos as it is a starting point.

\section*{Acknowledgments}
The author would like to acknowledge the support and useful comments of Professors Shin'ichi Oishi and Akira Koyama of Waseda University, and the helpful discussions of Professor Emeritus Yoshisuke Ueda of Kyoto University, Professor Naotsugu Chinen of the National Defense Academy of Japan, Dr. Kotaro Mine of the University of Tokyo, and particularly Professor Kazuhiro Kawamura of University of Tsukuba. This study was supported by the Japan Science and Technology Agency.

%\appendix

\end{document}